\documentclass[conference]{IEEEtran}

\IEEEoverridecommandlockouts

\usepackage{cite}
\usepackage{amsmath,amssymb,amsfonts}
\usepackage{algorithmicx}
\usepackage{graphicx}
\usepackage{textcomp}
\usepackage{xcolor}
\usepackage{subcaption}
\usepackage{comment}
\usepackage{caption}
\usepackage{tabularx}
\usepackage{array}
\usepackage{algorithm}
\usepackage{algpseudocode}
\usepackage{mdframed}

\def\BibTeX{{\rm B\kern-.05em{\sc i\kern-.025em b}\kern-.08em
    T\kern-.1667em\lower.7ex\hbox{E}\kern-.125emX}}

\begin{document}

\title{Secure and Decentralized Peer-to-Peer Energy Transactions using Blockchain Technology}

\author{Antar Kumar Biswas,~\IEEEmembership{Student Member,~IEEE,}
        Masoud H. Nazari,~\IEEEmembership{Senior Member,~IEEE}

\thanks{The authors are with the Department of Electrical and Computer Engineering, Wayne State University, Detroit, Michigan.
(e-mail: hr2122@wayne.edu; masoud.nazari@wayne.edu)}
}
\maketitle

\maketitle

\begin{abstract}

This paper presents an optimal peer-to-peer (P2P) energy transaction mechanism leveraging decentralized blockchain technology to enable a secure and scalable retail electricity market for the increasing penetration of distributed energy resources (DERs). A decentralized bidding strategy is proposed to maximize individual profits while collectively enhancing social welfare. The market design and transaction processes are simulated using the Ethereum testnet, demonstrating the blockchain network’s capability to ensure secure, transparent, and sustainable P2P energy trading among DER participants.

\end{abstract}


\section{Introduction}\label{sec:intro}

Blockchain is a technology that stores data in cryptographically linked blocks, creating a tamper-resistant, decentralized chain. It uses Distributed Ledger Technology (DLT) with consensus mechanisms and protocols to maintain ledger integrity through time-stamped, secured transaction blocks. The integration of blockchain with smart devices creates decentralized markets for resource and service sharing \cite{kaif2025blockchain,luo2025research}. This paper proposes an architecture that integrates blockchain and smart contracts for data integrity and transaction traceability, modeling decentralized energy transactions on the Ethereum testnet.

Over the past few years, decentralized energy transactions have become significant as a future strategy for smart community grids, allowing users to actively participate by buying, selling energy, or engaging in demand response (DR). 
For market clearing, various optimization methods can be employed, including 
the Alternating Direction Method of Multipliers \cite{aminlou2022peer, admm}, primal-dual gradient methods\cite{morstyn2018multiclass}, and consensus-based methods\cite{foo2024convergence} 
However, the integration of DERs into the present local electricity markets and incorporation of decentralized energy transactions can be challenging due to the complex and often centralized nature of traditional energy systems. 

In \cite{zhou2021credit,10454178}, authors investigate how blockchain technology can be used to support peer-to-peer (P2P) energy trading within DERs. A DR platform based on game theory has been developed to facilitate efficient and secure energy trading among prosumers \cite{9542944}.

However, existing studies often lack detailed transaction-level information about buyer and seller accounts within blockchain-based platforms. Additionally, critical transaction metadata—such as block hashes, block numbers, gas costs, and transaction statuses—is frequently underreported in the literature. This paper introduces a novel decentralized, blockchain-based energy trading mechanism tailored for retail electricity markets. The primary contributions of this work are:

1) The development of a P2P energy transaction architecture that determines optimal production levels for producers with the goals of minimizing cost, maximizing revenue, and enhancing social welfare.

2) The design and implementation of a detailed decentralized energy transaction framework on a blockchain platform, providing full transparency into transactional metadata and execution processes.

The paper is organized as follows: Section II introduces the network model and formulates the optimization problem. Section III presents the energy trading architecture. Section IV covers the experimental setup and results. Section V concludes and suggests future work.

\section{P2P Energy Market Modeling}
\subsection{Market Clearing Price }
The market clearing price (MCP) is the agreed-upon price at which energy is bought and sold, balancing supply and demand in the market. In this paper, MCP will be defined as the intersection point of the inverse demand and supply curve \cite{kirschen2018fundamentals}. First, the bid prices and the amount of trading energy from all buyers and ask prices from all sellers are collected. In this model, \text{inverse demand function is defined as:}
\[
D^{-1}(q) = 
\begin{cases}
\pi_{b1}, & 0 \leq q \leq q_{b1} \\
\pi_{b2}, & q_{b1} < q \leq q_{b1} + q_{b2} \\
\vdots & \\
\pi_{bn}, & \sum_{k=1}^{n-1} q_{bk} < q \leq \sum_{k=1}^{n} q_{bk}
\end{cases}
\]

\text{Also, the inverse supply function is defined as:}
\[
S^{-1}(q) = 
\begin{cases}
\pi_{s1}, & 0 \leq q \leq q_{s1} \\
\pi_{s2}, & q_{s1} < q \leq q_{s1} + q_{s2} \\
\vdots & \\
\pi_{sm}, & \sum_{l=1}^{m-1} q_{sl} < q \leq \sum_{l=1}^{m} q_{sl}
\end{cases}
\]

where $\mathcal{N} = \left \{ {1, 2, ..., n} \right \}$ and $\mathcal{M} = \left \{ {1, 2, ..., m} \right \}$ are the set of buyers and sellers respectively, MCP is represented as $\pi^*$, $\pi_{s}$ are the ask prices, and $\pi_b$ are the bid prices, $q_{bk}$ is the quantity that $n^{th}$ buyer wants to buy at price $\pi_{bn}$ and $q_{sl}$ is the quantity that $m^{th}$ seller is offering to sell at price $\pi_{sm}$.

\begin{figure} [h]
    \centering
    \includegraphics[width=0.45\textwidth]{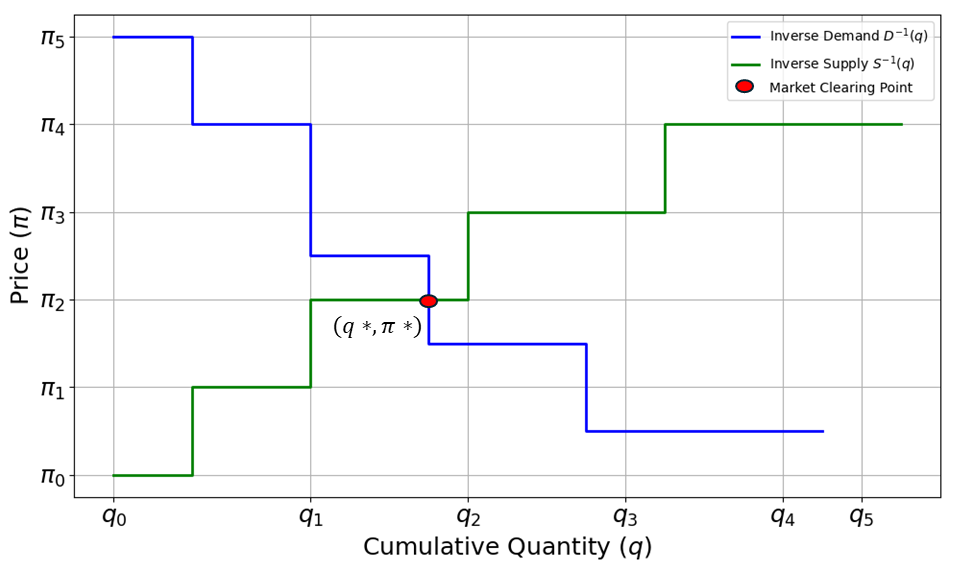}
    \caption{Supply and demand interaction.}
    \label{fig:MCP}
\end{figure}

Next, the intersection point of the inverse demand and supply is obtained. 
Fig. \ref{fig:MCP} shows the equilibrium condition in the energy market where the inverse demand curve intersects the inverse supply curve. This intersection determines the MCP ($\pi^*$) and the corresponding equilibrium quantity $(q^*)$, where supply meets the forecast demand.

\subsection{Social Welfare Function for Seller and Buyer}
\subsubsection{Cost Function of Producers}
The cost function $C_i(p_i)$ represents the cost incurred by producer $i$ to produce $p_i$ units of electricity. In this model, the cost function is typically quadratic, which can be expressed as \cite{powersystemanalysisbook}:

\begin{equation}
    C_i(p_i) = \alpha_ip_i^2+\beta_ip_i+\gamma_i
\end{equation}
where $\alpha_i$, $\beta_i$, and $\gamma_i$ are the three positive constant. 

\subsubsection{Utility Function}
The utility function $U_j(d_j)$ represents the satisfaction or benefit derived by consumer $j$ from consuming $d_j$ units of electricity. There are two types of utility functions commonly used for modeling electricity consumers: the logarithmic utility function \cite{liu2018energy} and the quadratic utility function \cite{doan2021peer}. In this model, the utility function is quadratic, which can be expressed as:

\begin{equation}\label{equ_6}
    U_j(d_j) = b_jd_j-\frac{\theta_j}{2}d_j^2
\end{equation}
where $b_j$ is the bid by the consumer $j$ and $\theta_j$ is the unwillingness of the consumer to consume $d_j$ amount of energy. For a higher value of $\theta_j$ the consumer is willing to reduce its demand, and the value of the utility function decreases with that.

\subsubsection{Objective Function}
The objective is to maximize social welfare ($SW$), which consists of consumer and producer surplus. Consumer surplus is the utility value minus the payment, while producer surplus is revenue minus production costs. To express mathematically \cite{khorasany2019decentralized},
\begin{equation}
\begin{matrix}
    \underset{p_i}{\max} \, \text{SW} = \sum_j^{n}(U_j(d_j)-\pi^* d_j)+ \sum_i^{m}(\pi^*p_i-C_i(p_i))
\end{matrix}
\end{equation}

This equation could be written as 
\begin{align}\label{equ.sw}
        \underset{p_i}{\max} \, \text{SW} = \sum_j^{n}(b_jd_j-\frac{\theta_j}{2}d_j^2-\pi ^*d_j)\nonumber \\
             +\sum_i^{m}(\pi^* p_i-\alpha_ip_i^2-\beta_ip_i-\gamma_i)
\end{align}

subject to,
    \begin{equation}\label{market equ}
    \sum_i^{n}p_i=\sum_j^{m}d_j=q^*
\end{equation}
\begin{equation}\label{non-negative}
    p_i\geq 0
\end{equation}

The market equilibrium constraint in (\ref{market equ}) indicates that total supply must meet the demand. Also, (\ref{non-negative}) ensures that the amount of production must be non-negative.

\subsection{Decentralized Optimization for Producers}

In decentralized optimization, the local subproblem refers to the task each participant (node, agent, prosumer, etc.) undertakes independently.
In this model, each producer has its own production cost function $C_i(p_i)$, which depends on the amount of production. The aim is to determine each producer's optimal production quantity $p_i$ in a decentralized manner, using an hour-ahead MCP ($\pi^*$) obtained from the intersection of the predicted inverse supply and demand curve. 

\subsubsection{Local Subproblem}
Each producer $i$ solves:
\begin{flalign}
    \mathcal{L}(p_i,\lambda) = \pi^* p_i-\alpha_ip_i^2 - \beta_ip_i - \gamma_i + \lambda p_i
\end{flalign}

Rearrange the equation as:
\begin{flalign}
    \mathcal{L}(p_i,\lambda) = (\pi^*+\lambda)p_i-\alpha_ip_i^2 - \beta_ip_i - \gamma_i
\end{flalign}
where, ($\alpha_ip_i^2+\beta_ip_i+\gamma_i$) is the individual cost function of production. Lagrange multiplier $\lambda$ is used in the decentralized optimization algorithm to iteratively enforce the supply-demand balance constraint. The term $\lambda p_i$ incentivizes or penalizes production based on the system's need to balance supply and demand. It mathematically reflects the incentive for producers to align their production with market equilibrium.

To find the optimal production $p_i$, we take the derivative of $\mathcal{L}_i$ with respect to $p_i$ to zero \cite{boyd2004convex}:
\begin{flalign}
    \frac{\delta\mathcal{L}_i}{\delta p_i}=  \pi^*-2\alpha_ip_i - \beta_i + \lambda =0
\end{flalign}

Then, we rearrange the equation to solve for $p_i$:
\begin{align}\label{opt pro}
    p_i=\frac{\pi^*-\beta_i+\lambda}{2\alpha_i}
\end{align}

This formula indicates that higher $\pi^*$ and $\lambda$ will increase $p_i$, while higher cost coefficients reduce the production.

\begin{mdframed}[leftline=false, rightline=false, innerleftmargin=0pt, innerrightmargin=0pt, innertopmargin=10pt, innerbottommargin=10pt]
\section*{Algorithm 1: Production Optimization Algorithm}

\begin{enumerate}
    \item \textbf{Initialize:} Given the predicted production, demand, and market clearing price. Set the initial $\lambda$.
    \item \textbf{Local Optimization:} Solve the local optimization for each producer $i$. Determine the optimal production level $p_i$ that minimizes the cost function adjusted by $\lambda$.
    \item \textbf{Total Production:} Calculate the total production by summing up $p_i$ from all producers.
    \item \textbf{Compute Imbalance:} Compute the imbalance of production and demand:
    \[
    \text{imbalance} = \sum_i p_i - \sum_j d_j
    \]
    \item \textbf{Update $\lambda$:} Update the Lagrange Multiplier $\lambda$:
    \[
    \lambda_{\text{new}} = \lambda_{\text{old}} + \text{step size} \times \text{imbalance}
    \]
    \item \textbf{Check for Convergence:} If the absolute value of the imbalance is less than the specified tolerance, stop the iteration:
    \[
    |\text{imbalance}| < \text{tolerance}
    \]
    \item \textbf{Determine Optimal Production:} Determine the optimal production $p_i^*$.
    \item \textbf{Calculate social welfare.}
\end{enumerate}
\end{mdframed}

\section{Blockchain-enabled Market Design}

\subsection{Secure Ethereum Network Implementation}

Blockchain technology is used to secure and validate energy transactions. Each transaction is recorded in a block, and once confirmed, it becomes part of an immutable ledger. Ethereum secures transactions using consensus mechanisms like Proof of Stake, cryptographic hashing, and smart contracts. These mechanisms validate transactions, ensure data immutability and traceability, and automatically execute agreements. Remix Ethereum Integrated Development Environment (IDE) is a web-based platform specifically designed for developing, testing, and deploying Ethereum smart contracts. Smart contracts are self-executing agreements deployed on Ethereum using the Solidity programming language. Remix IDE simplifies smart contract development with an integrated Solidity compiler, compiling contracts into Ethereum-compatible bytecode. Its debugging tools allow developers to execute and inspect contracts, pinpoint errors, and streamline deployment to testnets or local blockchains. The interaction and testing interface offers a user-friendly way to execute functions, monitor state changes, and analyze transactions.

\subsection{ Ethereum Trading Algorithm}

The MCP is computed off-chain using a social welfare optimization algorithm. This value was securely transmitted to the Ethereum testnet using a Chainlink oracle node, which then executed within the smart contract. This ensured synchronization between computational logic and on-chain enforcement.

\begin{figure}
    \centering
    \includegraphics[width=0.48\textwidth]{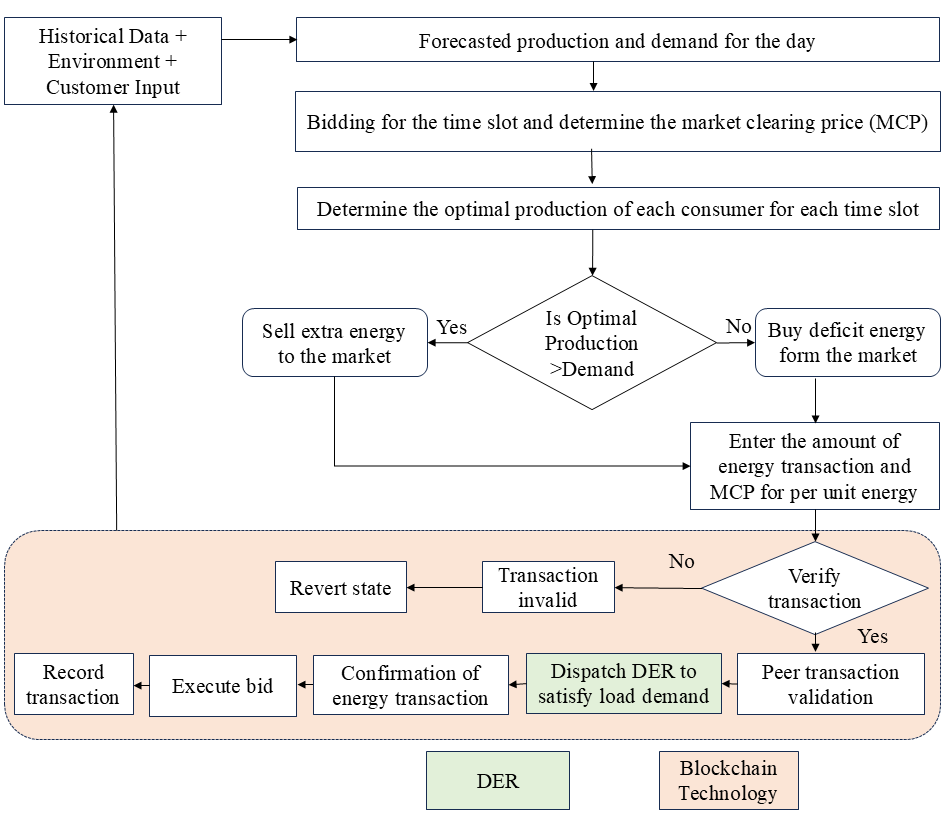}
    \caption{{ Blockchain market design and optimization algorithm.}}
    \label{fig:trading_strategy_flowchart}
\end{figure}

The detailed steps of the blockchain-enabled market is shown in Fig.~\ref{fig:trading_strategy_flowchart} and summarized as follows:   

\begin{itemize}
    \item \textbf{MCP Estimation}: Forecast demand and compute an initial MCP using historical data.
    
    \item \textbf{Optimization}: Run algorithm-1 to optimize the SW and allocate production across prosumers.
    
    \item \textbf{Buyer/Seller Role}: Determine each participant’s role—sell excess energy or buy to meet demand.
    
    \item \textbf{Chainlink Oracle}: Push the off-chain MCP and optimal production result to the smart contract securely.
    
    \item \textbf{Remix IDE / Smart Contract}: Execute energy transactions (buy/sell) on the Ethereum testnet using Solidity.
    
    \item \textbf{Blockchain Record}: Automatically record and finalize the transaction on the blockchain.
    
    \item \textbf{Smart Meter Validation}: Validate energy usage to confirm actual trade amounts before Ether is transferred.
\end{itemize}

When transaction is confirmed, a unique cryptographic hash is generated. Each block contains its own hash along with the hash of the preceding block, forming a secure, tamper-resistant blockchain.

\section{Simulation Results} 

\begin{figure}[h]
  \centering
  \includegraphics[width=0.35\textwidth]{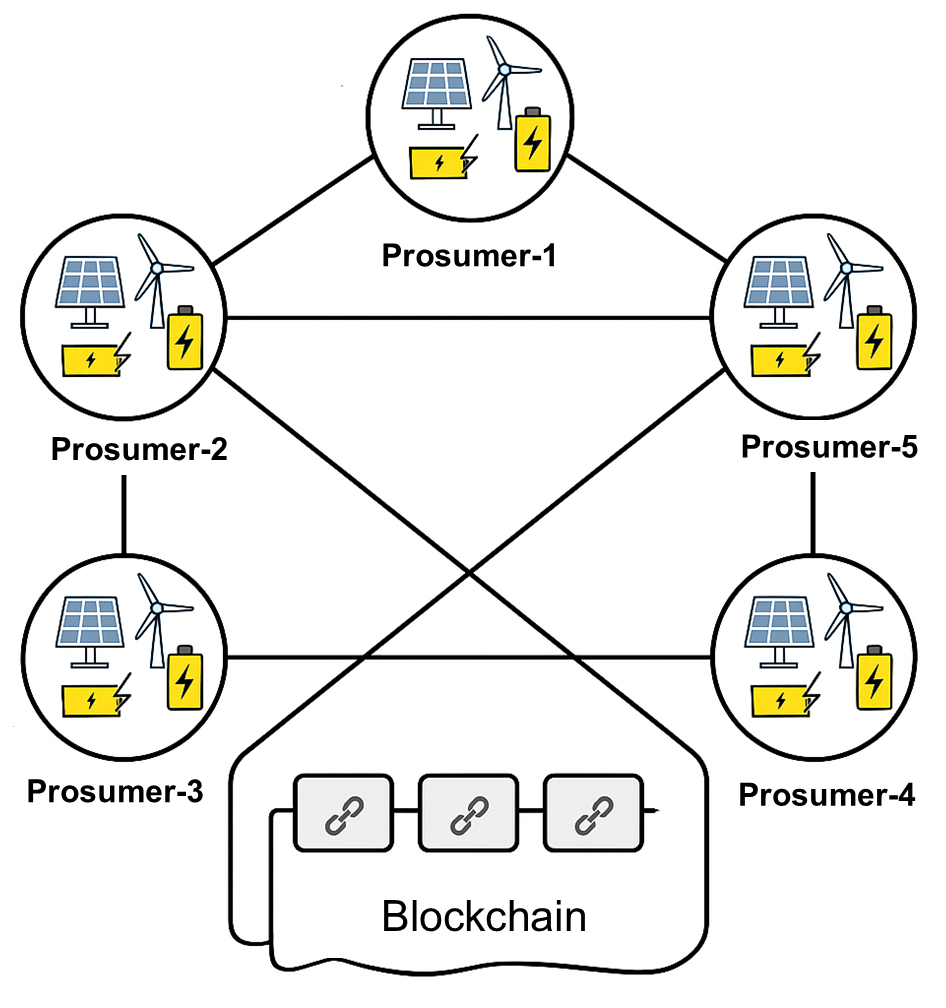}
  \captionsetup{font=small, justification=centering}
  \caption{ 5-Prosumer network diagram.}
  \label{fig:5-prosumer}
\end{figure}

\begin{figure}[h]
  \centering
  \includegraphics[width=0.45\textwidth]{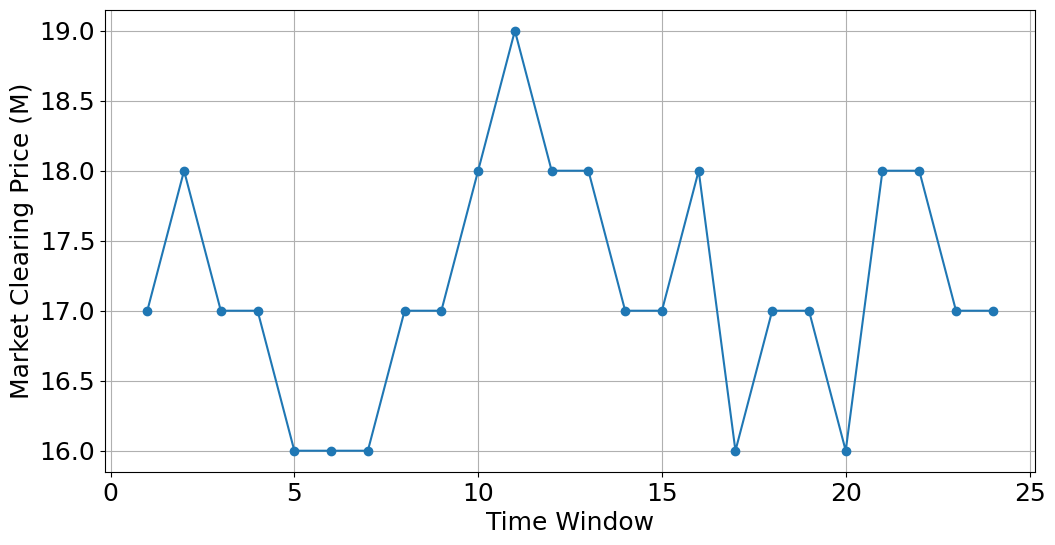}
  \captionsetup{font=small, justification=centering}
  \caption{Market Clearing Price for 24-hours time window.}
  \label{fig:MCP day}
\end{figure}

A network comprising five prosumers was considered for the simulation study (Fig. \ref{fig:5-prosumer}). Each prosumer is equipped with distributed energy resources (DERs) and battery energy storage systems (BESS), enabling them to generate and store electricity. A prosumer operates as a producer when its generation exceeds demand, and as a consumer when its demand surpasses generation. The optimal electricity production for each prosumer was determined using a social welfare maximization algorithm (\ref{equ.sw}). Simulations were executed on a system with an Intel Core i7 processor (2.60 GHz) and 16 GB of RAM.
Figure \ref{fig:MCP day} illustrates the hourly market clearing price (MCP) schedule over a 24-hour period, where each time window represents the MCP for one hour.

\begin{figure*}[h]
    \centering
    \includegraphics[width=0.8\textwidth]{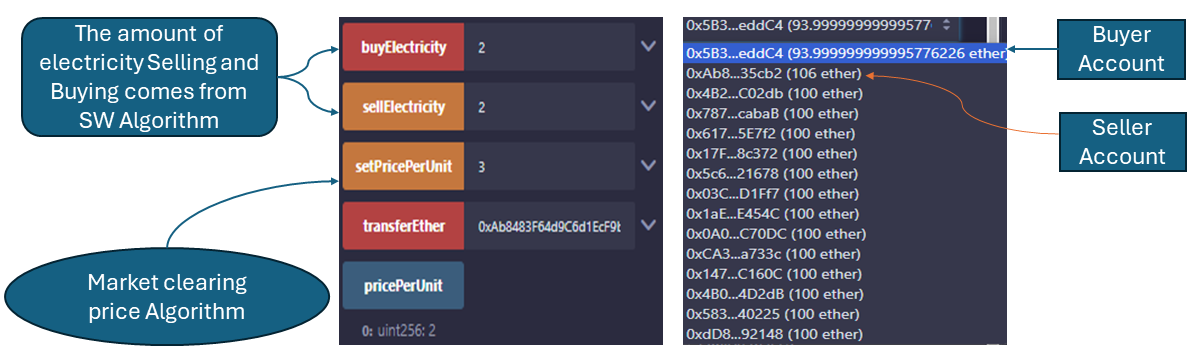}
    \caption{{ Transaction using a blockchain platform.}}
    \label{fig: connection blockchain and market algorithm}
\end{figure*}

\begin{figure}[h]
  \centering
  \includegraphics[width=0.45\textwidth]{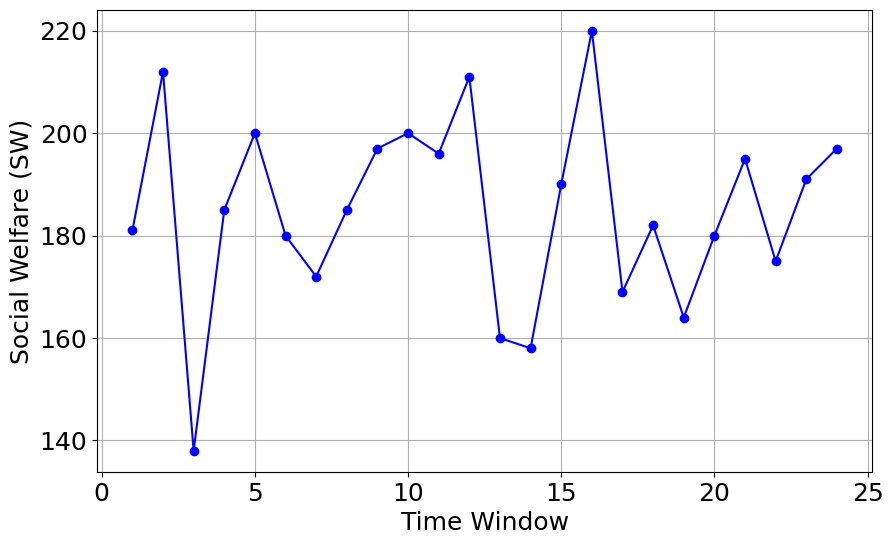}
  \captionsetup{font=small, justification=centering}
  \caption{Social Welfare for 24-hour time window.}
  \label{fig: SW}
\end{figure}

There is a positive correlation between social welfare and production values. This indicates that social welfare is higher when the production levels are closer to their optimal values. If actual production deviates significantly from these optimal levels, social welfare is likely to decrease due to inefficiencies in the market. An increase in MCP is associated with an increase in social welfare. Fig. \ref{fig: SW} shows the social welfare value for different time windows.

\begin{table}[ht]
\centering 
\caption{Parameters value of Cost and Utility Function}
\label{tab:utility_costs}
\begin{tabular}{|p{1.3cm}|p{1.3cm}|p{1.3cm}|p{1.3cm}|p{1.3cm}|}
\hline
\textbf{Prosumer} & \textbf{Alpha($\alpha$)} & \textbf{Beta($\beta$)} & \textbf{Gamma($\gamma$)} & \textbf{Theta($\theta$)} \\
\hline
1 & 1.0 & 2 & 3 & 0.5 \\
\hline
2 & 1.1 & 2.5 & 3.5 & 0.2 \\
\hline
3 & 0.9 & 1 & 4 & 0.25 \\
\hline
4 & 1.2 & 1.5 & 2 & 0.3 \\
\hline
5 & 1.3 & 1 & 5 & 0.45 \\
\hline
\end{tabular}
\end{table}

\begin{figure}[h]
  \centering
  \includegraphics[width=0.48\textwidth]{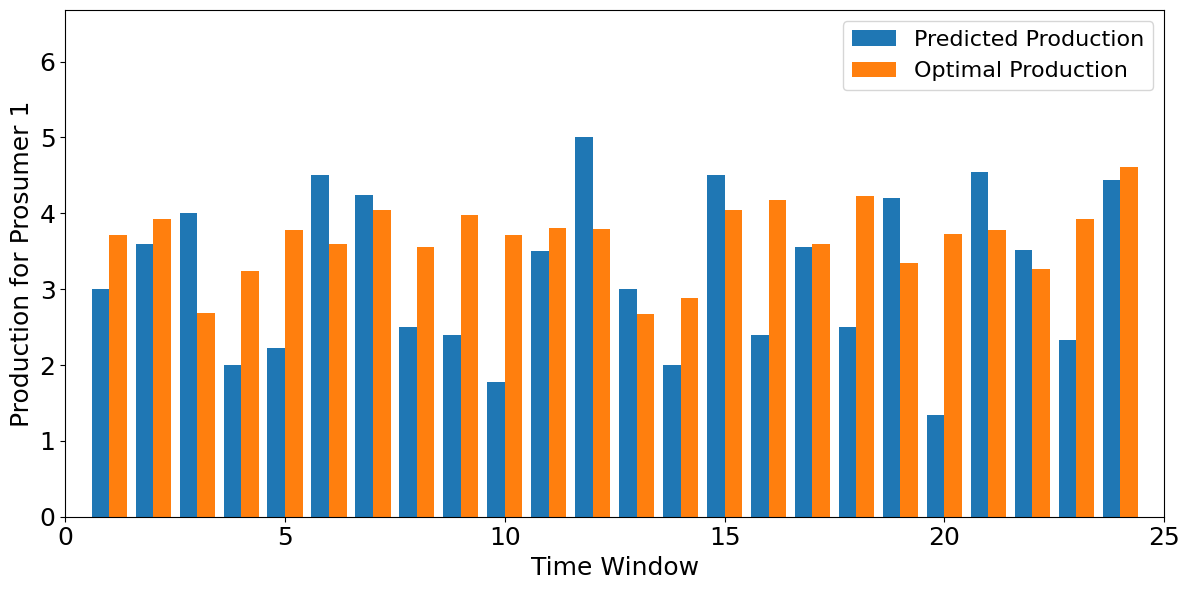}
  \captionsetup{font=small, justification=centering}
  \caption{Predicted vs Optimal production for Prosumer 1.}
  \label{fig: prosumer 1}
\end{figure}

Using the formula in (\ref{opt pro}) and Algorithm 1, the optimal individual output for each prosumer is computed, where the predicted production is estimated based on day-ahead generation plans. Figure \ref{fig: prosumer 1} shows the comparison between the predicted and optimal energy production for Prosumer 1. Similar comparisons were carried out for the remaining four prosumers to evaluate and improve the efficiency of the overall energy transaction process.

A key aspect of the paper involves conducting energy trading using blockchain technology. The Ethereum IDE is utilized to facilitate energy transactions. Fig. \ref{fig: connection blockchain and market algorithm} details a practical execution scenario of the implemented smart contract. Electricity trades are initiated through smart contract functions.
The blockchain transaction flow for this study is as follow:
\begin{itemize}
    \item Input MCP into smart contract via ``setPricePerUnit".
    \item Smart contract functions ``buyElectricity" and ``sellElectricity" are used to input optimal electricity quantities for buyers and sellers based on maximizing social welfare.
    \item Using ``transferEther", execute financial transactions automatically, directly linked to energy transactions.
    \item Ethereum nodes validate the transactions, including verification of sufficient funds, appropriate function calls, and adherence to smart contract conditions.
\end{itemize}

In Fig. \ref{fig: connection blockchain and market algorithm}, the required information for the transaction is provided such as the amount of energy trading, MCP for the time window, and buyer and seller account details. After the transaction, the blockchain platform transfers Ethereum from buyer to seller. For instance, in Fig. \ref{fig: connection blockchain and market algorithm}, 2 p.u of electricity are transferred from the seller’s account to the buyer’s account. Given that the unit price is set at 3 Ether, a total of 6 Ether is transferred from the buyer to the seller as payment. 
The transaction details are recorded in Table \ref{tab:transactions}.

The blockchain transaction algorithm is detailed as follows:
\begin{itemize}
    \item ``0x1 Transaction mined and execution succeeded" implies that the transaction has been successfully validated, included in a block, and executed without errors.
    \item The transaction hash (0x2cb8d457344a69919f60ab80ae4 6fc00e34eb35c0ea22524a4f3803d396289c2) allows any user or auditor to trace, verify, and review the transaction.
    \item The recorded block hash (0x3f97c74c39d8e77f73fda 4b051e11faea7e8a1455d7a35ede60679411dd3d64) ensures that the transaction is in the blockchain.
    \item Block number 11 indicates this specific transaction's position within the blockchain's sequential ledger, allowing easy retrieval and verification.
    \item 0x583b0a6a701c58f545d0cfc803fc88f75f6bedc4 is the buyer's Ethereum address from which Ether was sent to pay for purchased electricity.
    \item ElectricityMarket.transferEther(address)0xdada34b0bf16f 56196a5e41ab18776a75482d is the seller's address, which receives funds.
    \item A gas limit of 38465 was set for this transaction, indicating the computational capacity allowed for execution. Note that, gas measures the computational effort needed for the execution of a transaction.
    \item Actual gas used was 33447 for the transaction. 
    \item An execution cost of 12015 gas is needed to run the logic and data changes of the smart contract.
\end{itemize}

\begin{table}[ht]
\centering 
\caption{Transaction Details of a P2P Trading}
\label{tab:transactions}
\begin{tabular}{|p{2cm}|p{6cm}|}
\hline
\textbf{Field} & \textbf{Details} \\
\hline
Status & 0x1 Transaction mined and execution succeeded \\
\hline
Transaction Hash & \texttt{0x2cb8d457344a69919f60ab80ae46f c00e34eb35c0ea22524a4f3803d3962 89c2} \\
\hline
Block Hash & \texttt{0x3f97c74c39d8e77f73fda4b 051e11faea7e8a1455d7a35ede6067 9411dd3d64}  \\
\hline
Block Number & 11 \\
\hline
From & \texttt{0x583b0a6a701c58f545d0cfc803fc 88f75f6bedc4} \\
\hline
To & \texttt{ElectricityMarket.transferEther (address) 0xdada34b0bf16f56196a5 e41ab18776a75482d} \\
\hline
Gas & 38465 gas \\ \hline
Transaction Cost & 33447 gas \\ \hline
Execution Cost & 12015 gas \\ \hline
\end{tabular}
\end{table}


\section{Conclusion and Future Directions}

This paper has explored the potential of blockchain technology in decentralized energy transactions, emphasizing its capacity to enhance the transparency and security of electricity markets. Through the implementation and analysis of an optimized bidding system within a blockchain-based framework, this paper has demonstrated significant advancements in meeting the multiple objectives of maximizing social welfare, streamlining energy distribution among distributed prosumers, and ensuring secure energy transactions. The application of the model on the Ethereum testnet has provided valuable insights into its practical implementation and effectiveness in a controlled environment.

Future work should focus on new trading mechanisms by incorporating supervised and machine learning for market trend predictions, enabling bi-directional vehicle-to-grid energy flows, integrating IoT for optimized energy storage communication, and facilitating inter-community trading with grid services. Additionally, examine the benefits and impacts of DER and P2P trading on the distribution grid.

\bibliographystyle{ieeetr}
\bibliography{main}

\end{document}